\documentstyle[aps,amssymb]{revtex}
\def\openone{\leavevmode\hbox{\small1\kern-3.8pt\normalsize1}}
\def\<{\langle} \def\>{\rangle}%
\def\grp{{\mathbf G}}
\def\Sn{{\mathbf S}_N}

\def\Ccal{{\cal C}}

\def\Lcal{{\cal L}}

\def\Rnmbr{{\mathbb R}}
\def\Ecal{{\cal E}}

\def\Kcal{{\cal K}}
\def\Ical{{\cal I}}
\def\Hcal{{\cal H}}
\def\Mcal{{\cal M}}
\def\Tr{\hbox{Tr}}
\def\dim{\hbox{dim}}

\begin{document}
\title{Optimal Non-Universally Covariant Cloning}
\author{G. M. D'Ariano, P. Lo Presti}
\address{Theoretical Quantum
Optics Group\\ Universit\`a degli Studi di Pavia and INFM Unit\`a di
Pavia\\ via A. Bassi 6, I-27100 Pavia, Italy}
\date{\today}
\twocolumn \maketitle
\begin{abstract}
We consider non-universal cloning maps, namely cloning transformations 
which are covariant under a proper subgroup $\grp$ of the universal unitary
group $U(d)$, where $d$ is the dimension of the Hilbert space $\Hcal$ 
of the system to be cloned. We give a general method for optimizing 
cloning for any cost-function. Examples of
applications are given for the phase-covariant cloning (cloning of equatorial
qubits) and for the Weyl-Heisenberg group (cloning of ``continuous variables'').
\end{abstract}

\section{Introduction}
The impossibility of perfectly cloning an unknown input state is a typical
quantum feature \cite{no-cloning},
nonetheless, in the laws of quantum mechanics there's enough room 
either to systematically produce approximate copies \cite{BuzekHillery}
or to make perfect copies of orthogonal states \cite{Yuen} or of 
non-orthogonal ones with a non-unit probablility \cite{DuanGuo}.
These possibilities have been studied in several works 
\cite{GisinMassar,Werner,BrussEkertMacchiavello}.

Recently, quantum cloning has entered the realm of experimental physics
\cite{SimonWeihsZeilinger}\cite{parametric gates}. Moreover it has became
interesting from a pratical point of view, since it can be used to speed-up
some quantum computations \cite{GalvaoHardy} or to
perform some quantum measurements \cite{clon and meas,BrussCalsamigliaLutkenhaus}.
All these tasks require a spreading of the quantum information contained 
in a system into a larger system, and quantum cloning is a way to achieve
such a spreading.

In this paper we will see how any ``spreading'' corresponds
to a particular completely-positive (CP) map.
By exploiting the correspondence between CP-maps and positive operators
on the tensor product of the output and input spaces \cite{Jamiolkowski},
we can parametrize all the possible spreading transformation.
Then we focus on covariant CP-map, showing that quantum cloning 
is a particular case of permutation-covariance. By means of Schur's lemmas
we completely characterize the positive operators corresponding to quantum
cloning transformations. By the same technique, we characterize 
$\grp$-covariant cloning transformation, where $\grp$ is any single-copy 
covariance group.

The parametrization of CP-maps, and in particular of cloning and
covariant cloning, stands at the base of any further optimization.
In fact, quantum cloning can be used to perform some tasks on the
copies, and depending on what these copies will be used for, one defines
a ``goodness'' criterion for the cloning process and 
optimizes accordingly.

The paper is organized as follows.
In Sec. \ref{sec: cloning}, we briefly describe a quantum cloning
transformation and its relation to CP-maps. Sec. \ref{sec: CP and op} is
devoted to the description of CP-maps in terms of positive operators,
while in Sec. \ref{sec: cov CP} we treat the case of covariant CP-maps, giving
their parametrization with suitable covariant positive operators.
In Sec. \ref{sec: clon opt} we use the previously explained techniques
to deal with cloning optimization, focusing on the covariant case. 

\section{Cloning transformations}
\label{sec: cloning}
In a quantum cloning transformation, 
the input state $\rho\in\Lcal(\Hcal)$
is processed in order to
produce $N$ output clones (throughout the paper $\Lcal(\Hcal)$ will denote
the vector space of linear bounded operators on the Hilbert space $\Hcal$).
This requires a ``spreading'' of 
$\rho$ into the joint state $\rho'\in\Lcal(\Hcal^{\otimes N})$ of $N$ 
identical quantum systems.
The most general setup for such purpose is the following.
Initially, $\rho$ is encoded in a quantum system $S_1$, while $N-1$ equivalent
systems $S_i$, $i=2\ldots N$, are prepared in a fixed state $|\omega\>_{(N-1)}$.
An auxiliary system $E$ is provided in a state $|e\>$, in order to make
the whole system isolated.
A unitary transformation $U$ acts on the overall state producing 
the output
\begin{equation}
\Lambda = U\,\rho\otimes(|\omega\>
\<\omega|)_{(N-1)}\otimes|e\>\<e|\,U^\dagger\;.
\label{sigma}
\end{equation}
By taking the partial trace of $\Lambda$ on the auxiliary system, we get
the joint state $\rho'$ of the $N$ output
systems $S_i$. This state will eventually support
the clones. Upon calculating the trace with respect to a chosen 
basis $\{|j\>_E\}$ for $\Hcal_E$ one has
\begin{equation}
 \rho'=\sum_{j=1}^{dim\Hcal_E}\, _E\<j|\Lambda|j\>_E=
\sum_{j=1}^{dim\Hcal_E} A_j \rho A^\dagger_j\doteq\Ecal(\rho)\;,
\label{rho'}
\end{equation}
where
$A_j=\,_E\<j|U|\omega\>_{(N-1)} |e\>_E$.

The map $\rho\rightarrow\Ecal(\rho)$ 
in Eq. (\ref{rho'}) is a completely-positive (CP) and 
trace-preserving linear map from $\Lcal(\Hcal)$ to $\Lcal(\Hcal^{\otimes N})$.
Trace preserving CP-maps generally describe the evolution of open quantum systems.
To understand the general features of quantum cloning and for sake of optimization,
it is convenient to treat these 
maps at an abstract level: a realization theorem guarantees that any CP-map
can be achieved as a unitary transformation on an extended 
Hilbert space \cite{Kraus,Ozawa}, similarly to Eq. (\ref{sigma}).
CP-maps will be shortly reviewed in the next session.

\section{CP-maps and positive operators}
\label{sec: CP and op}
A linear map $\Ecal:\Lcal(\Hcal)\rightarrow\Lcal(\Kcal)$ is 
completely-positive if its trivial extension $\Ecal\otimes\Ical_{\Hcal'}$ 
to $\Lcal(\Hcal\otimes\Hcal')$ is positive, for any $\Hcal'$ ($\Ical_{\Hcal'}$
denoting the trivial map on $\Lcal(\Hcal')$).

Here we recall a convenient notation \cite{bellobs}.
Fixing two orthonormal basis $\{|i\>_1\}$ and $\{|j\>_2\}$ for $\Hcal_1$
and $\Hcal_2$ respectively, any vector $ |\Psi\>\!\>\in\Hcal_1\otimes\Hcal_2$ can
be written as
\begin{equation}
|\Psi\>\!\>=\sum_{ij}c_{ij} |i\>_1|j\>_2 \doteq |C\>\!\>\;,
\label{kket}
\end{equation}
where $C=\sum_{ij}c_{ij} |i\>_{1\,2}\<j|\,\in\Lcal(\Hcal_2,\Hcal_1)$ is a linear bounded operator from $\Hcal_2$ to $\Hcal_1$.
\\The following relations can be easily verified
\begin{eqnarray}
&A\otimes B |C\>\!\>= |ACB^T\>\!\> \;,\label{kket formulas} \\
&\Tr_{\Hcal_2}\big[|A\>\!\>\<\!\<B|\big]
=AB^\dagger\,\in\Lcal(\Hcal_1)\;.&
\label{part trace}
\end{eqnarray}

For every CP-map $\Ecal: \Lcal(\Hcal) \rightarrow \Lcal(\Kcal)$
we define the positive operator $R_\Ecal$ in $\Lcal(\Kcal\otimes\Hcal)$
\begin{equation}
R_\Ecal \doteq \Ecal\otimes\Ical\,\big(|\openone\>\!\>\<\!\<\openone|\big)\;,
\label{R}
\end{equation}
where $\Ical$ denotes the identical map over the extention space $\Hcal$,
and for the vector $|\openone\>\!\>\in\Hcal\otimes\Hcal$ 
we used the notation (\ref{kket}) for $\Psi=\openone$ the identity 
matrix with respect to a fixed basis on $\Hcal$.
The action of $\Ecal$ on $\rho \in \Lcal(\Hcal)$ can be expressed as
\begin{equation}
\Ecal(\rho)=\Tr_\Hcal\big[\,\openone\otimes\rho^T\;R_\Ecal\big]\;,
\label{E via R}
\end{equation}
where the transposition $\rho\rightarrow\rho^T$
is performed with respect to the same fixed basis.
In fact, substituting Eq. (\ref{R}) in Eq. (\ref{E via R}), one has
\[\Tr_{\Hcal}\big[\,\openone\otimes \rho^T\; R_\Ecal\,\big]
=\Tr_{\Hcal}\big[\,\openone\otimes \rho^T\; \Ecal\otimes\Ical
\big(|\openone\>\!\>\<\!\<\openone|\big)\,\big]\;.\]
Then, it is possible to take the factor $\openone\otimes \rho^T$ 
inside the CP-map
$\Ecal\otimes\Ical$, since they act independently on different spaces.
By applying Eq. (\ref{kket formulas}), one obtains
\[\Tr_{\Hcal}\big[\,\openone\otimes \rho^T\; R_\Ecal\,\big]
=\Tr_{\Hcal}\big[\,\Ecal\otimes\Ical
\big(|\rho\>\!\>\<\!\<\openone|\big)\,\big]\;,\]
and thus, commuting the partial trace with $\Ecal\otimes\Ical$,
and using Eq. (\ref{part trace}), one finally gets 
Eq. (\ref{E via R}).

The operator $R_\Ecal$ is the only one for
which Eq. (\ref{E via R}) holds true. In fact, suppose $R_\Ecal$ and
$R'$ give the same CP-map $\Ecal$ by means of Eq. (\ref{E via R}), then
\[ \Tr_{\Hcal}\big[\,\openone\otimes\rho^T\;
(\,R_\Ecal-R'\,)\,\big]=0\in\Lcal(\Kcal)\;,\quad \forall
\rho\in\Lcal(\Hcal)\;.\] 
Since an operator $O\in\Lcal(\Hcal\otimes\Kcal)$
is null if $\<v|O|v\>=0\in\Lcal(\Kcal)$ for all $|v\>\in\Hcal$,
it follows that $R_\Ecal=R'$.
Thus, the correspondence from CP-maps to 
positive operators is ``into''.

Since $R_\Ecal$ is positive, it can be written as
\begin{equation}
R_\Ecal=\sum_i |A_i\>\!\>\<\!\<A_i| \;,
\label{diag R}
\end{equation}
where there are many different choices of the vectors $|A_i\>\!\>$, which
are not necessarily eigenvectors of $R_\Ecal$, and generally are 
not normalized.

Substituting this relation in Eq. (\ref{E via R}) and remembering 
that $A_i\in\Lcal(\Hcal,\Kcal)$, we find
\begin{equation}
\Ecal(\rho)=\sum_i \Tr_\Hcal\big[\,\openone\otimes\rho^T\;
|A_i\>\!\>\<\!\<A_i| \,\big]=\sum_i \, A_i \,\rho\, A_i^\dagger\;,
\label{Kraus}
\end{equation}
thus recovering that any CP-map admits different Kraus's decompositions \cite{Kraus},
depending on the choice of the vectors $|A_i\>\!\>$ in Eq. (\ref{diag R}).\\
Clearly, Eq. (\ref{diag R}) holds for any positive
operator $R$ on $\Kcal\otimes\Hcal$. The map defined by 
$R$ through Eq. (\ref{E via R}) is completely-positive,
since it can be expressed in the form of Eq. (\ref{Kraus}) which
trivially gives a CP-map. Then the correspondence from CP-maps to operators
is also ``onto''.

Concluding, Eq. (\ref{E via R}) defines a one-to-one correspondence
between CP-maps from $\Lcal(\Hcal)$ to $\Lcal(\Kcal)$ and positive 
operators on $\Kcal\otimes\Hcal$. By exploiting this correspondence,
properties of $\Ecal$ can be translated into properties of $R_\Ecal$.
For example, the trace-preserving condition for $\Ecal$
\[
\Tr_\Kcal[\,\Ecal(\rho)\,]=1
=\Tr_\Hcal\big[\,\rho^T\,\Tr_\Kcal[\,R_\Ecal\,]\,\big]\;,\]
for all $\rho\in\Lcal(\Hcal)$ such that $\Tr[\rho]=1$, becomes
\begin{equation}
\Tr_\Kcal[\,R_\Ecal\,]=\openone\in\Lcal(\Hcal)\;.
\label{trace pres}
\end{equation}

In the following, it will be useful to consider the dual map $\Ecal^\vee$
of a CP-map $\Ecal$, namely
the transformation in the Heisenberg picture versus the Schr\"oedinger picture
map $\rho\rightarrow\Ecal(\rho)$. The dual map $\Ecal^\vee$ is defined by the
identity
\begin{equation}
\Tr\big[\,\rho\,\Ecal^\vee(O)\,\big]=\Tr\big[\,\Ecal(\rho)\,O\,\big]\;,
\end{equation}
which must be valid for all operators $O\in\Lcal(\Kcal)$.
In terms of the operator $R_\Ecal$ one has
\begin{equation}
\Ecal^\vee(O)=\Tr_\Kcal\big[\,O\otimes\openone\,R_\Ecal^{T_\Hcal}\,]\;,
\end{equation}
where $T_\Hcal$ denotes partial transposition on the Hilbert space $\Hcal$ only \cite{T non CP}.

In the next session, the correspondence $\Ecal\leftrightarrow R_\Ecal$ 
will be applied to the covariance
condition for a CP-map, which turns out to be the key idea to deal with
cloning and covariant cloning.

\section{Covariant CP-maps}
\label{sec: cov CP}
Let $\Ecal: \Lcal(\Hcal) \rightarrow \Lcal(\Kcal)$ be a CP-map, and let
$\grp$ be a group with unitary representations $U$ and $V$ on $\Hcal$ and
$\Kcal$ respectively. $\Ecal$ is $\grp$-covariant with respect to $U$ and $V$
if 
\begin{equation} 
\Ecal\big(U_g\,\rho\,U_g^\dagger\big)=V_g\,\Ecal(\rho)\,V_g^\dagger\;,
\label{cov1} 
\end{equation}
for any $\rho\in\Lcal(\Hcal)$ and $g\in\grp$.\\
By means of Eq. (\ref{E via R}), the covariance condition becomes
\begin{eqnarray}
\Ecal(\rho)&=&\Tr_\Hcal\big[\,\openone\otimes\rho^T\;R_\Ecal\,\big]=
\nonumber\\
&\equiv&\Tr_\Hcal\big[\,\openone\otimes\rho^T\;V_g^\dagger\otimes U_g^T\;
R_\Ecal\;V_g\otimes U_g^*\,\big]\;.
\label{cov2}
\end{eqnarray}
From the uniqueness of the operator associated to a CP-map, we
conclude that $\Ecal$ is $\grp$-covariant if and only if
\begin{equation}
R_\Ecal=V_g^\dagger\otimes U_g^T\;
R_\Ecal\;V_g\otimes U_g^*\;,\quad\forall g\in\grp\;,
\label{cov3}
\end{equation}
or equivalently
\begin{equation}
\big[\,R_\Ecal\,,\,V_g\otimes U_g^*\,\big]=0\;,\quad\forall g\in\grp\;.
\label{cov}
\end{equation}
Thus, $\grp$-covariance of a CP-map $\Ecal$ is equivalent to 
\hbox{$\grp$-invariance} of the corresponding positive operator $R_\Ecal$.

\subsection*{Group invariant operators}
\label{subsec: inv op}
The $\grp$-representation $W={\mathrm diag}(V\otimes U^*)$ 
on $\Kcal\otimes\Hcal$, defined as $W_g=V_g\otimes U^*_g$,
is generally reducible, i. e. the space can be decomposed into a direct sum
of minimal invariant subspaces $\Mcal_i$
\begin{equation}
\Kcal\otimes\Hcal=\bigoplus_{i=1}\Mcal_i\;,
\label{space decomp}
\end{equation}
each $\Mcal_i$ supporting a unitary irreducible representation (UIR)
of the group.
Given this decomposition one can look at any operator $O$ on 
$\Kcal\otimes\Hcal$ as a set of operators $O_j^i$ in $\Lcal(\Mcal_j,\Mcal_i)$,
so that $O=\sum_{ij}O^i_j$.

Due to irreducibility of the subspaces $\Mcal_i$, $W_g$ will be decomposed as follows
\[ (W_g)_j^i = \delta_{ij}\,T^{j}_g \;, \]
where $T^j$ is the UIR supported by $\Mcal_j$. Two UIR $T^i$ and $T^j$ are
equivalent, $i\!\sim\! j$, if they are connected by similarity,
i. e. through an isomorfism $I^i_j\in\Lcal(\Mcal_j,\Mcal_i)$ such that 
$T^j=(I^i_j)^{\scriptstyle -1} T^i I^i_j$.

The invariance equation (\ref{cov3}) becomes
\[ T^i_g  \,R^i_j \, T^j_g = R^i_j\,\quad \forall g\in\grp\;,\]
so that, by Schur's lemmas (see, for example, Ref. \cite{Jones}), 
one finally has
\begin{equation}
R^i_j =c_{ij} \,I^i_j \;,
\label{def c}
\end{equation}
where if $i\!\nsim\! j$ then $c_{ij}=0$, and, if $i\!\sim \! j$,
$c_{ij}$ can be different from zero.

Since equivalent representations are related by similarity, 
in any invariant subspace $\Mcal_i$ one can choose
the basis $\{|i,l\>, \,l=1\ldots\mbox{\small dim}\Mcal_i\}$
so that for $i\!\sim\! j$
\begin{equation}
\<i,l| T^i |i,m\>=\<j,l| T^j |j,m\> \;,
\label{base def}
\end{equation}
hence
\begin{equation}
I^i_j =\sum_l  |i,l\>\<j,l|\doteq\openone^i_j\;,
\label{I_j^i def}
\end{equation}
and finally
\begin{equation}
R=\sum_{ij}c_{ij}\openone^i_j\;.
\label{cov R}
\end{equation}
In order to have a positive $R$, the matrix $c_{ij}$ must be positive,
since taking $|\psi\>\!\>=\sum_i\sum_{l=1}^{\scriptscriptstyle{\mathrm
dim}\Mcal_i}\psi_{il}\,|i,l\>$ one has
\[ \<\!\<\psi|R|\psi\>\!\>=\sum_{ij}
   \sum_{l=1}^{\scriptscriptstyle{\mathrm dim}\Mcal_i}
   \psi_{li}^* \,c_{ij} \,\psi_{lj}\;. \]
Recalling that $c_{ij}=0$ if $i\nsim j$, and reordering the indices
of the representations by grouping the equivalent ones,
the matrix $c_{ij}$ assumes a block diagonal form,
different blocks corresponding to inequivalent representations, each
block including all representations equivalent to the same one.
In this way, each block has dimension equal 
to the multiplicity of the representation.
Positivity of $R$ implies positivity of each block of matrix $c_{ij}$.
This structure of $c_{ij}$ is reflected on $R$ by means of Eq. (\ref{cov R}).

\section{Optimal covariant cloning}
\label{sec: clon opt}
A cloning map is just a CP-map $\Ccal$ 
from $\Lcal(\Hcal)$ to $\Lcal(\Hcal^{\otimes N})$
with the output copies invariant under the permutations of the $N$
output spaces.
This is equivalent to a particular covariance of the CP-map $\Ccal$ 
for the group of permutations $\Sn$, namely it corresponds
to the invariance of the positive operator $R_\Ccal$
under the representation $W={\mathrm diag}(V\otimes I)$, 
where $V$ is the representation of 
$\Sn$ permuting the $N$ identical output spaces, and $I$ (corresponding
to $U$ in Eq. (\ref{cov1})) is the $\Sn$-trivial 
representation on the input space.
One has
\begin{equation} 
\Ccal(\rho)=V_\pi\,\Ccal(\rho)\,V^\dagger_\pi\;,\quad \forall\,\pi\in\Sn\;.
\label{clon eq}
\end{equation}
Notice that permutation covariance does not imply that the output state has
support in the symmetric subspace of the output space $\Hcal^{\otimes N}$.

As explained in the previous section, Eq. (\ref{clon eq})
determines a peculiar block structure for the operator $R_\Ccal$
associated to the map $\Ccal$. Such a structure 
is strictly related to the decomposition of $\Hcal^{\otimes (N+1)}$
into invariant subspaces for $V_\pi\otimes\openone$. Any possible
cloner is described by an $R_\Ccal$ with that structure and satisfying 
the trace-preserving condition of Eq. (\ref{trace pres}).
In this way, one classifies all possible cloning maps through the decomposition
into irreducibles of the $\Sn$-representation $V$ on $\Hcal^{\otimes N}$.

In addition to permutation invariance, in this paper we will consider
covariance under a group of transformations $\grp$, with 
representation $T$ on $\Hcal$.
This corresponds to the following identity
\begin{equation}
\Ccal(T_g\,\rho\,T_g^\dagger)=T_g^{\otimes N}\,\Ccal(\rho)\,
T_g^{\dagger\otimes N}\;.
\label{grp cov}
\end{equation}

One can choose a cost function $C(R_\Ccal)$ (related to 
the following usage of the clones).
Covariant cloning is suited to invariant cost-functions of the form
\begin{equation}
C(R_\Ccal)\equiv C_0(\rho_0, R_\Ccal)=C_0(T_g\rho_0 T_g^\dagger, R_\Ccal)\;,
\end{equation}
where $\rho_0$ is the seed of the covariant family of states on which
we are interested in having the minimum $C_0$.

The best cloner is found by minimizing $C(R_\Ccal)$ vs $R_\Ccal$,
with the constraints of positivity $R_\Ccal\geq 0$, trace-preserving 
(Eq. (\ref{trace pres})), and covariance under permutations and $\grp$
(Eqs. (\ref{clon eq}) and (\ref{grp cov})).

\section{Examples}
\subsection*{Phase covariant qubit cloning}
Here, we consider the problem of cloning
a qubit  in a $U(1)$-covariant fashion,
where the group representation is given by
\begin{equation}
T_\phi= \exp\left[\frac i 2\, \phi\,(\openone-\sigma_z)\right]\;.
\end{equation}
Since the cloning to two copies is already given in Ref. \cite{macchiav},
whereas the general case for $N$ copies is very complicated, here for 
simplicity we will consider the case of $N=3$ copies.
We want to achieve the maximum fidelity between input and clones, when the 
input is an equatorial qubit
\begin{equation}
|\psi_\phi\>=T_\phi \frac{1}{\sqrt 2}[|0\>+|1\>]=
\frac{1}{\sqrt 2}[|0\>+e^{i\phi}|1\>]\;.
\end{equation}
In other terms, we want to maximize the fidelity
\begin{eqnarray}
F_\phi&=&\Tr
\big[\,\Ccal(|\psi_\phi\>\<\psi_\phi|)\,|\psi_\phi\>\<\psi_\phi|\,\big]
\equiv F_0=\nonumber\\
&=&\Tr\big[\,\openone^{\otimes 2}\otimes|\psi_0\>\<\psi_0|
\otimes (|\psi_0\>\<\psi_0|)^T \,R_\Ccal\,\big]\;.
\label{eq fidelity}
\end{eqnarray}
Since the equator is invariant even for spin flipping,  
 here we will require the additional covariance 
with respect to the group ${\mathbb Z}_2$, 
with representation $\{\openone, \sigma_x\}$.

In order to satisfy all the covariance requirements, $R_\Ccal$ must be 
invariant for permutations, phase shift, and spin flip,
i. e. for products of any of the following unitary operators
\[ T_\phi^{\otimes 3}\otimes T_\phi^*\;,\quad 
  V_\pi\otimes\openone\;,\quad
  \sigma_x^{\otimes 3}\otimes \sigma_x^*\;. \]

The Hilbert space $\Hcal^{\otimes 3+1}$ can be decomposed into subspaces
which are irreducible with respect to the joint action of $U(1)$ and ${\bf S}_3$,
as shown in Table \ref{table}.

\begin{table}[h]
\begin{center}
\begin{tabular}{|c|c|c|c|c|}
Space &  Unnormalized Basis &  $U(1)$ & ${\bf S}_3$& Flipped\\
\hline \hline 
$\Mcal_1$ & $|0001\>$ & -1 & T & $\Mcal_5$\\ \hline
$\Mcal_2$ & $|0000\>$ &  0 & T & $\Mcal_6$\\ \hline
$\Mcal_3$ & $|1001\>+|0101\>+|0011\>$ & 0 & T & $\Mcal_7$\\ \hline
$\Mcal_4$ & 
$\begin{array}{c}
|1001\>-|0101\>\;,\\ \frac12|1001\>+\frac12|0101\>-|0011\>
\end{array}$ & 0 & D & $\Mcal_8$ \\ \hline
$\Mcal_5$ & $|1110\>$ & 3 & T & $\Mcal_1$\\ \hline
$\Mcal_6$ & $|1111\>$ & 2 & T & $\Mcal_2$\\ \hline
$\Mcal_7$ & $|0110\>+|1010\>+|1100\>$ & 2 & T & $\Mcal_3$\\ \hline
$\Mcal_8$ & 
$\begin{array}{c}
|0110\>-|1010\>\;,\\ \frac12|0110\>+\frac12|1010\>-|1100\>
\end{array}$ & 2 & D & $\Mcal_4$\\ \hline
$\Mcal_9$ & $|1000\>+|0100\>+|0010\>$ & 1 & T & $\Mcal_{10}$\\ \hline
$\Mcal_{10}$ & $|0111\>+|1011\>+|1101\>$ & 1 & T  & $\Mcal_{9}$\\ \hline
$\Mcal_{11}$ & 
$\begin{array}{c}
|1000\>-|0100\>\;,\\ \frac12|1000\>+\frac12|0100\>-|0010\>
\end{array}$ & 1 & D  & $\Mcal_{12}$\\ \hline
$\Mcal_{12}$ &
$\begin{array}{c}
|0111\>-|1011\>\;,\\ \frac12|0111\>+\frac12|1011\>-|1101\>
\end{array}$ & 1 & D & $\Mcal_{11}$
\end{tabular}
\caption{$\Hcal^{\otimes 3+1}$ decomposition into 
$U(1)-{\bf S}_3$ irreducibles.
$U(1)$ acts on each subspace as a phase shift $e^{i n\phi}$, where
$n\in{\mathbb Z}$ (column III) labels inequivalent representation.
${\bf S}_3$ acts trivially (T) on one-dimensional subspaces, 
whereas on bidimensional ones it acts as the defining representation (D).
Spin flipping connects subspaces (column V).
\label{table} }
\end{center}
\end{table}
Looking at Table \ref{table}, one can see that in this example
the matrix $c_{ij}$ defined in Sec. \ref{sec: cov CP}
has the following positive diagonal blocks:
\[$\{1\}$,\; $\{2,3\}$,\; $\{4\}$,\; $\{5\}$,\; $\{6,7\}$,\; 
$\{8\}$, \;$\{9,10\}$, \; $\{11,12\}$.\]
To ensure spin flipping covariance, 
the elements of $c_{ij}$ connected by a flip
must be equal, for example $c_{23}=c_{67}$.

At the end, to fill the blocks of $c_{ij}$ in the right way,
we need the parameters
$a,\;b,\;c,\;d,\;e,\;f,\;g\;\in\Rnmbr^+$, ${\bf v}\in\Rnmbr^3$, where
$d\geq e$, $f\geq g$, and $c\geq|\!|{\bf v}|\!|$.
Table \ref{table2} explains how to employ them.
\begin{table}[h]
\begin{center}
\begin{tabular}{|c|c|}
Blocks& Content\\ \hline\hline
$\{1\}$, $\{5\}$&$a$\\ \hline
$\{4\}$, $\{8\}$&$b$\\ \hline
$\{2,3\}$, $\{6,7\}$&$c\,\openone+{\mathbf v \cdot \sigma}$\\ \hline
$\{9,10\}$&$d\,\openone+e\,\sigma_x$
\\ \hline
$\{11,12\}$&$f\,\openone+g\,\sigma_x$
\end{tabular}
\caption{Content of the blocks of the matrix $c_{ij}$, chosen in order to have
$R_\Ccal$ describing the most general CP-map from $\Lcal(\Hcal)$
to $\Lcal(\Hcal^{\otimes 3})$ which is covariant with respect to
permutations, phase shift, and spin flip.
\label{table2} }
\end{center}
\end{table}

The parameters must satisfy another constraint given by the 
trace-preserving condition defined in Eq. (\ref{trace pres}). 
Within this parametrization it reads
\begin{equation}
a+2b+2c+d+2f=1\;.
\label{ph:tr pres}
\end{equation}
Substituting this equation into the equatorial fidelity $F_0$ 
defined in Eq. (\ref{eq fidelity}), one has
\begin{equation}
F_0=\frac12+\frac13(e-g)+\frac{\sqrt3}{3}v_x\;.
\label{eq fidelity2}
\end{equation}
This quantity can be easily maximized by hand, taking into account 
the constraint given by Eq. (\ref{ph:tr pres}) and the properties of
the parameters. The maximum fidelity is $F=\frac56$ (see note \cite{note})
and is achieved for $d=e=1$ and all the other parameters equal to zero.
The optimal phase covariant cloning is thus described by the operator
\begin{equation}
R_\Ccal^{opt}=|\Phi\>\!\>\<\!\<\Phi|\;,
\end{equation}
where
\begin{eqnarray}
|\Phi\>\!\>=&{\scriptstyle\frac{1}{\sqrt3}}[&|1000\>+|0100\>+|0010\>
+\nonumber\\ &+&|0111\>+|1011\>+|1101\>]\;.\nonumber
\end{eqnarray}
The Kraus's decomposition of the optimal cloner is
$\Ccal(\rho)=\,B\,\rho\,B^\dagger$, where
\begin{eqnarray}
 B=&{\scriptstyle\frac1{\sqrt3}}[&|{100}\>\<{0}|
+|{010}\>\<{0}|+|{001}\>\<{0}|+\nonumber\\
&+&|{011}\>\<{1}|+|{101}\>\<{1}|+|{110}\>\<{1}|]\;.\nonumber
\end{eqnarray}
\subsection*{Cloning of continuous variables}
The parametrization of CP-maps given in Sec. \ref{sec: CP and op}
and its specialization to the covariant case are useful tools for
engineering measurements. The idea is to ``spread'' a quantum
state on a larger system with a CP-map $\Ecal$, and then 
to perform a measurement on the spread state.
The connection between the POVM $\{M_i\}$ 
on the larger space $\Kcal$
and the resulting one $\{M^\vee_i\}$ on $\Hcal$ is given by
\begin{equation}
M^\vee_i=\Ecal^\vee(M_i)\doteq\Tr_\Kcal\big[\,M_i\otimes\openone\;
R_\Ecal^{T_\Hcal}\,\big]\;,
\label{dual CP}
\end{equation}
where $\Ecal^\vee$ is the dual map of $\Ecal$, and the symbol
$T_\Hcal$ stands for transposition with respect to $\Hcal$ only 
(see Sec. \ref{sec: CP and op}).

In Ref. \cite{clon and meas}, the cloning map for continuous variables
of  Ref. \cite{cerf} is used to achieve the optimal POVM 
for the joint measurement
of two conjugated quadratures $X_0$ and $X_{\frac\pi 2}$
of an oscillator mode $a$ 
(where $X_\phi={\small \frac12}[a^\dagger e^{i\phi}+a e^{-i\phi}]$)
by measuring them separately on the two clones.
Here, we will briefly show how our general method works on this problem.

Denote by $\Hcal_3$ the input space and by $\Hcal_1$, $\Hcal_2$ 
the two output spaces of the oscillator modes $a_3$, $a_1$, $a_2$ respectively.
The cloning is described by
\begin{equation}
R_\Ccal=\frac12P_{12}\otimes\openone_3\;
	\openone_1\otimes(|\openone\>\!\>\!\<\!\<\openone|)_{23}\;
	P_{12}\otimes\openone_3\;,
\label{R cerf}
\end{equation}
where $P=V\,|0\>\!\<0|\!\otimes\!\openone \,V^\dagger$, and
$V$ is the 50\% beam splitter unitary transformation 
$V=\exp[\frac\pi 4(a_1^\dagger a_2-a_1 a_2^\dagger)]$.

A simple calculation shows that
\begin{equation}
P=\frac2\pi\int \hbox{d}^2\alpha\, |\alpha\>\<\alpha|\,^{\otimes 2}\;,
\label{P}
\end{equation}
where $|\alpha\>=D(\alpha)|0\>$, and 
$D(\alpha)\!=\!e^{\alpha a^\dagger -\bar\alpha a}$ 
is the displacement operator generating the {\em Weyl-Heisenberg} ($WH$) group.
By means of Eq. (\ref{P}), the invariance of $R_\Ccal$ defined
in Eq. (\ref{R cerf}) with respect to permutations and displacements
can be easily verified.

Using the dual cloning map as in Eq. (\ref{dual CP}), we should check that
\begin{equation}
\Ccal^\vee(E_x^0\otimes E_y^{\pi/2})
=\frac1\pi|\alpha\>\<\alpha|\;,\quad\alpha=x+iy\;,
\label{coh POVM}
\end{equation}
where
$E^\phi_v\!=\!|v\>_\phi \,_\phi\<v|$, and $X_\phi|v\>_\phi\!=\!v|v\>_\phi$.
In fact, the last term of Eq. (\ref{coh POVM}) is the
well-known optimal POVM for the joint measurement of conjugated quadratures,
whereas $E^\phi_v$ is the POVM of the $\phi$-quadrature measurement.
Hence identity (\ref{coh POVM}) guarantees that the cloning achieves 
the optimal joint measurement of the two conjugated quadrature via 
commuting measurements on clones.

Noticing that
\begin{equation}
E_x^0\otimes E_y^{\pi/2}=D(\alpha)^{\otimes2} E_0^0\otimes E_0^{\pi/2}
D(\alpha)^{\otimes2\dagger}\;,
\end{equation}
and exploiting the $WH$ covariance, Eq. (\ref{coh POVM}) reduces to
\begin{equation}
\Ccal^\vee(E_0^0\otimes E_0^{\pi/2})=\frac1\pi|0\>\<0|\;.
\end{equation}
Substituting Eq. (\ref{dual CP}) into this last equation,
and taking matrix elements $\<i|\ldots|j\>$,
one finally must check that
\begin{equation}
\,_0\<0|\,_{\frac{\pi}{2}}\<0|\<i|R_\Ccal|0\>_0|0\>_{\frac{\pi}{2}}|j\>=
\frac1\pi\delta_{i0}\,\delta_{j0}\;.
\label{vac prj}
\end{equation}
Since $V|0\>_0|0\>_{\frac{\pi}{2}}=\sqrt{\frac2\pi}|\openone\>\!\>$ 
and $V|0\>|0\>=|0\>|0\>$ (see Ref. \cite{mauromax}),
one has that $P|0\>_{\frac{\pi}{2}}|0\>_0=\sqrt{\frac2\pi}|0\>|0\>$. Thus Eq.
(\ref{vac prj}) holds, and the cloning really achieves the wanted POVM.

\subsection*{Universal cloning}
Clearly, the universal covariant cloning of Werner \cite{Werner}
is a special case of covariant cloning for the covariance 
group $U(d)$, $d=\hbox{dim}\Hcal$, of all unitary operators
on $\Hcal$. 
Here, for sake of comparison to Ref. \cite{Werner}, we consider more
generally the cloning from $M$ to $N>M$ copies. Hence the cloning is a
CP-map $\Ccal$ from
$\Lcal(\Hcal^{\otimes M})$ to $\Lcal(\Hcal^{\otimes N})$
such that for any $U\in U(d)$ and $\sigma\in\Lcal(\Hcal)$
\begin{equation}
\Ccal(U^{\otimes M}\sigma^{\otimes M}U^{\dagger\otimes M})
 =U^{\otimes N}\Ccal(\sigma^{\otimes M})U^{\dagger\otimes N}\;.
\end{equation}
The cost-function for optimization 
is the (negative) fidelity between clones and input
\begin{equation}
C(R_\Ccal)=-F=-\Tr\big[\,\sigma^{\otimes N}\,
\Ccal(\sigma^{\otimes M})\,\big]\;,
\end{equation}
where $\sigma$ is pure.
Owing to covariance, the fidelity $F$ does not
depend on $\sigma$, since any pure state lies in the $U(d)$ orbit 
of any other pure state.

The optimal cloning map of Ref. \cite{Werner} is given by
\begin{equation}
\Ccal(\rho)=\frac{d(M)}{d(N)}S_N (\rho\otimes\openone^{\otimes (N-M)})S_N\;,
\end{equation}
where $\rho\in\Lcal(\Hcal^{\otimes M})$, $S_N$ is 
the projector on the symmetric subspace $\Hcal^{\otimes N}_+$,
and $d(N)=\dim(\Hcal^{\otimes N}_+)$.
In our framework, one has
\begin{equation}
R_\Ccal=\frac{d(M)}{d(N)}\,\tilde S\,\openone_{\Hcal^{\otimes (N-M)}}
(|\openone\>\!\>\<\!\<\openone|)_{\Hcal^{\otimes (M+M)}}\,\tilde S\;,
\end{equation}
where $\tilde S= S_N \otimes \openone^{\otimes M}$. It can be easily 
verified that $R_\Ccal$ is both covariant and permutation invariant
as it must be.

\section*{Acknowledgements}
This work has been supported by the Italian Ministero
dell'Universit\`a e della Ricerca Scientifica e Tecnologica (MURST)
under the co-sponsored project 1999 {\em Quantum Information
Transmission And Processing: Quantum Teleportation And Error
Correction}. 


\begin{thebibliography}{}
\bibitem{T non CP} Remember that transposition is not a CP-map, hence
partial transposition is not a positive map. Thus the partial transposition
of a positive operator can be non-positive.
\bibitem{note} We notice that this fidelity value is larger than
the bound given in Ref. \cite{macchiav}. The correctness of the 
present value and the optimality of the map can be checked a posteriori.
\end{thebibliography}
\end{document}